\documentclass{article}
\usepackage{graphicx} 

\usepackage{graphicx}
\usepackage{amsmath}
\usepackage{amsfonts,amssymb}
\usepackage[version=4]{mhchem}
\usepackage[english]{babel}
\usepackage[colorlinks,bookmarks=false,citecolor=black,linkcolor=blue,urlcolor=blue]{hyperref}
\usepackage[T1]{fontenc}
\usepackage[utf8]{inputenc}
\usepackage{dsfont}
\usepackage{caption}
\usepackage{subcaption}
\usepackage{tikz-feynman}
\usepackage{wrapfig}
\usepackage{enumitem}

\usepackage{parskip} 

\usepackage{authblk}
\usepackage{bbold}
\usepackage{mathtools}
\usepackage{tcolorbox}
\usepackage{accents}
\usepackage{braket}
\usepackage{physics}
\usepackage[onehalfspacing]{setspace}
\usepackage{slashed}
\usepackage{pgfplots}
\pgfplotsset{compat=newest}
\pgfplotsset{plot coordinates/math parser=false}
\newlength\figureheight
\newlength\figurewidth 
\usepackage{float}
\usepackage[left=2cm,right=2cm,top=2cm,bottom=1.5cm]{geometry}

\title{Early-career researchers perspective on future colliders} 
\date{\today}

\tiny

\author[1]{Julia~Allen}
\author[2]{Bruno~Alves}
\author[*,3]{Jan-Hendrik~Arling}
\author[4]{Kamil~Augsten}
\author[*,5]{Emanuele~Bagnaschi}
\author[7]{Anna~Bennecke}
\author[8]{Cecilia~Borca}
\author[9]{Paulo~Braz}
\author[*,10]{Lydia~Brenner}
\author[10]{Jordy~Degens}
\author[9]{Yannick~Dengler}
\author[11]{Christina~Dimitriadi}
\author[12]{Eleonora~Diociaiuti}
\author[53]{Patrick~Dougan}
\author[5]{Laurent~Dufour}
\author[13]{Patrick~Dunne}
\author[14]{Ozgur~Etisken}
\author[5]{Silvia~Ferrario~Ravasio}
\author[15]{Nikolai~Fomin}
\author[10]{Andrea~Garcia~Alonso}
\author[16]{Leif~Gellersen}
\author[17]{Andreas~Gsponer}
\author[4]{Tomas~Herman}
\author[18]{Bojan~Hiti}
\author[19]{Laura~Huhta}
\author[*,20]{Armin~Ilg}
\author[21]{Kateřina~Jarkovská}
\author[22]{Jelena~Jovicevic}
\author[23]{Lucia~Keszeghova}
\author[24]{Henning~Kirschenmann}
\author[11]{Suzanne~Klaver}
\author[18]{Arman~Korajac}
\author[25]{Anastasia~Kotsokechagia}
\author[26]{Meike~Kussner}
\author[27]{Aleksandra~Lelek}
\author[28]{Guiseppe~Lospalluto}
\author[29]{Péter~Major}
\author[22]{Veljko~Maksimovic}
\author[30]{Jakub~Malczewski}
\author[31]{Carla~Marin~Benito}
\author[32]{Paula~Martinez~Suarez}
\author[33]{Vukasin~Milosevic}
\author[34]{Atanu~Modak}
\author[35]{Arnau~Morancho~Tarda}
\author[36]{Laura~Moreno~Valero}
\author[37]{Elisabeth~Niel}
\author[5]{Nikiforos~Nikiforou}
\author[18]{Anja~Novosel}
\author[19]{Petja~Paakkinen}
\author[*,38]{Holly~Pacey}
\author[39]{Rute~Pedro}
\author[*,+,20]{Marko~Pesut}
\author[25]{Guillaume~Pietrzyk}
\author[40]{Michael~Pitt}
\author[41]{Vlad-Mihai~Placinta}
\author[42]{Archita~Rani~Dash}
\author[17]{Géraldine~Räuber}
\author[43]{Mariana~Shopova}
\author[44]{Radoslav~Someonov}
\author[45]{Sinem~Simsek}
\author[46]{Kirill~Skovpen}
\author[47]{Filomena~Sopkova}
\author[39]{Fernando~Souza}
\author[*,48]{Elisabetta~Spadaro~Norella}
\author[*,49]{Marta~Urbaniak}
\author[50]{Lourdes~Urda~Gomez}
\author[16]{Erik~Wallin}
\author[51]{Valentina~Zaccolo}
\author[5]{Nima~Zardoshti}
\author[52]{Grzegorz~Zarnecki}

\affil[+]{Editor}
\affil[*]{Future colliders WG}

\affil[1]{University of Edinburgh, Edinburgh; United Kingdom}
\affil[2]{Ecole Polytechnique, Ile-de-France; France}
\affil[3]{Deutsches Elektronen-Synchrotron DESY, Hamburg; Germany}
\affil[4]{Faculty of Nuclear Sciences and Physical Engineering, Czech Technical University in Prague, Prague; Czech Republic}
\affil[5]{CERN, Geneva; Switzerland}
\affil[6]{INFN Laboratori Nazionali del Gran Sasso, L'Aquila; Italy}
\affil[7]{Université Catholique de Louvain, Brussels; Belgium}
\affil[8]{University of Torino and INFN, Turin; Italy}
\affil[9]{Institute for Physics, University of Graz, Graz; Austria}
\affil[10]{Nikhef National Institute for Subatomic Physics and University of Amsterdam, Amsterdam; Netherlands}
\affil[11]{Uppsala University, Uppsala; Sweden}
\affil[12]{National Laboratory of Frascati and INFN, Frascati; Italy}
\affil[13]{Blackett Laboratory, Imperial College London, London; United Kingdom}
\affil[14]{Kirikkale University, Kirikkale; Turkey}
\affil[15]{University of Bergen, Bergen; Norway}
\affil[16]{Lund University, Lund; Sweden}
\affil[17]{Institute of High Energy Physics, Austrian Academy of Sciences, Vienna; Austria}
\affil[18]{Department of Experimental Particle Physics, Jožef Stefan Institute and Department of Physics, University of Ljubljana, Ljubljana; Slovenia}
\affil[19]{Department of Physics, University of Jyväskylä and Helsinki Institute of Physics, University of Helsinki, Helsinki; Finland}
\affil[20]{Physik-Institut, University of Zürich, Zürich; Switzerland}
\affil[21]{Faculty of Mathematics and Physics, Charles University, Prague; Czech Republic}
\affil[22]{Institute of Physics, University of Belgrade, Belgrade; Serbia}
\affil[23]{Faculty of Mathematics, Physics and Informatics, Comenius University, Bratislava; Slovakia}
\affil[24]{Department of Physics, University of Helsinki, Helsinki; Finland}
\affil[25]{IRFU, CEA, Université Paris-Saclay, Gif-sur-Yvette; France}
\affil[26]{Institut für Experimentalphysik, Ruhr-Universität Bochum; Germany}
\affil[27]{Universiteit Antwerpen, Antwerpen; Belgium}
\affil[28]{Institute for Particle Physics and Astrophysics, ETH Zürich, Zürich; Switzerland}
\affil[29]{MTA-ELTE Lendület CMS Particle and Nuclear Physics Group, Eötvös Loránd University, Budapest; Hungary}
\affil[30]{Henryk Niewodniczanski Institute of Nuclear Physics Polish Academy of Sciences, Kraków; Poland}
\affil[31]{Institut de Ciencies del Cosmos, Universidad de Barcelona, Barcelona; Spain}
\affil[32]{Institut de Fisica d'Altes Energies, Universidad Autonoma Barcelona, Barcelona; Spain}
\affil[33]{Institute of High Energy Physics, Beijing; China}
\affil[34]{Rutherford Appleton Laboratory, Oxford; United Kingdom}
\affil[35]{Niels Bohr Institute, University of Copenhagen, Copenhagen; Denmark}
\affil[36]{Institut für Theoretische Physik, Westfälische Wilhelms-Universität Münster, Münster; Germany}
\affil[37]{École Polytechnique Fédérale de Lausanne, Lausanne; Switzerland}
\affil[38]{University of Oxford, Oxford; United Kingdom}
\affil[39]{Laboratório de Instrumentação e F\'isica Experimental de Part\'iculas - LIP, Lisboa; Portugal}
\affil[40]{Department of Physics, Ben-Gurion University of the Negev, Beer-Sheva; Israel}
\affil[41]{Horia Hulubei National Institute of Physics and Nuclear Engineering, Bucharest-Magurele; Romania}
\affil[42]{Westfälische Wilhelms-Universität Münster, Münster; Germany}
\affil[43]{Institute for Nuclear Research and Nuclear Energy, Bulgarian Academy of Sciences, Sofia; Bulgaria}
\affil[44]{Sofia University, Sofia; Bulgaria}
\affil[45]{Istinye University, Istanbul; Turkey}
\affil[46]{Ghent University, Ghent; Belgium}
\affil[47]{Slovak Academy of Sciences, Bratislava; Slovakia}
\affil[48]{University of Genoa and INFN, Genoa; Italy}
\affil[49]{University of Silesia in Katowice, Katowice; Poland}
\affil[50]{Centro de Investigaciones Energ\'eticas Medioambientales y Tecnol\'ogicas (CIEMAT), Madrid; Spain}
\affil[51]{University of Trieste and INFN, Trieste; Italy}
\affil[52]{Henryk Niewodniczanski Institute of Nuclear Physics Polish Academy of Sciences, Kraków; Poland}
\affil[53]{The University of Manchester, Manchester; United Kingdom} 

\makeatletter         
\renewcommand\maketitle{
{
\begin{center}
{\huge \setstretch{1.2}\@title \par}
\vspace{1cm}
{\large \textbf{The ECFA Early Career Researcher's (ECR) Panel}}
\\[1cm]
{\large \@date}
\\[1cm]
\begin{minipage}{0.82\textwidth}
\normalsize Since its inception, the Large Hadron Collider (LHC) has significantly advanced particle physics and will continue to do so in the context of the High Luminosity LHC (HL-LHC) program to collect $3000$ fb$^{-1}$ by the end of 2041. The particle physics community worldwide is discussing which future collider could follow in the footsteps of the LHC and uncover yet inaccessible phenomena.

To foster the discussion on this important topic among the young particle physicist community, the Early-Career Researchers (ECR) panel of the European Committee for Future Colliders (ECFA) has organized the \textit{Future Colliders for Early-Career Researchers} workshop at CERN in September 2023. This document aims to summarise this event and present the ECR perspective, outline the key questions that came up during the discussions, and explore how ECRs can influence the decision process of future colliders community and beyond.  
\end{minipage}
\end{center}
\vspace{1.5cm}
\begin{flushleft}
{The ECFA Early Career Researcher's (ECR) Panel: \href{mailto:ecfa-ecr-organisers@cern.ch}{ecfa-ecr-organisers@cern.ch}\\[0.5cm]\@author}
\end{flushleft}}}
\makeatother

\begin{document}
\maketitle
\clearpage

\section{Introduction}
To understand the future trajectory of particle physics, it is essential to draw insights from the important milestones of the past century that provide guidance and valuable lessons for future endeavours. From the development of Quantum Physics and the theory of Relativity in the early 1900s to the discovery of the heaviest particles, the detection of neutrino oscillations, and the observation of the Higgs boson, the field of particle physics has been an extremely successful scientific undertaking of the past century that considerably extended the boundaries of human knowledge. These discoveries were made possible by sophisticated tools and techniques: detectors, colliders, accelerators, theoretical calculations, and computational models. These tools allowed physicists to reveal and understand the hidden structures at the smallest possible scales. Their conception, design, development, and construction also significantly impacted and shaped the emergence of new technologies, materials, and techniques with important societal impact and world-wide collaboration. As it has been made clear in the past hundred years, the evolution and the discoveries in particle physics as well as the collaborative efforts to construct and operate ever more sophisticated colliders and experiments extend well beyond their scientific scopes, yielding not only valuable scientific results but also practical applications in various technological domains and many societal benefits.

Exploring New Physics (NP) and pursuing a more comprehensive and precise understanding of Standard Model (SM) processes necessitate new machines with improved capabilities compared to existing colliders. In 2020, the last update of the European Strategy for Particle Physics \cite{European_Strategy_Update2020} was approved by the CERN council. One of its twenty strategy statements states that an electron-positron Higgs factory is the highest priority next collider. Considering the long timelines that these projects have, it is of paramount importance for early career researchers to participate in an informed way in the numerous discussions currently taking place on the future of the field. Any long-term decision related to building a new High Energy Physics (HEP) machine must transcend scientific and technological considerations. Hence, it is crucial to understand not only all the scientific ramifications of such a massive project but also to assess the socio-economic constraints and sustainability issues. This document aims to provide a comprehensive review of key considerations and issues to be taken into account when building the next collider, from an Early-Career Researcher (ECR) point of view. Its content summarises a one-day workshop organized at CERN in September 2023\footnote{\href{https://indico.cern.ch/event/1293507/}{Future Colliders for Early-Career Researchers}}, whose main objectives were to provide ECRs with an introduction to the prospective collider projects currently undergoing consideration so that young researchers can form their opinions and get actively involved in this important matter for the future of the field, and to foster the discussion within the ECR community on the same topic. The result of the workshop was a list of questions that ECRs think need to be answered in the context of a future collider: 
\begin{enumerate}[left=0.4cm,font=\bfseries]
    \itemsep-0.5em 
    \item What are the physics questions we wish to be answered?
    \item How can we make sure that the physics program is diverse enough?
    \item Are several smaller colliders preferable over one large collider for the diversity of the achieved physics program?
    \item What are the possibilities for the upgrades of the proposed projects?
    \item Taking realistic improvements in theory predictions into account, how precise can we get?
    \item How can we ensure collaboration with experiments probing other energy ranges ?
    \item Are the future collider timelines compatible with ECR careers given possibly long time gaps after HL-LHC?
    \item Would, and could, muon colliders be developed in time to follow the HL-LHC?
    \item Can the gap between HL-LHC and a large future collider be bridged with enough attractive projects?
    \item How can we ensure that a next collider is sustainable in terms of energy use?
    \item At what time scale should the ECR community dedicate itself to one particular proposal?
    \item How can ECRs make the impact they desire on the decision-making process?
\end{enumerate}

In the following document, some of these questions will be addressed, starting with question 1.

\section{Why do we need a new HEP machine?}

The field of particle physics stands today at a unique point in its history. After over a century of exploration, involving the discovery of new particles and the validation of numerous predictions made by Quantum Field Theory (QFT) and the Standard Model (SM), the field has undeniably provided significant knowledge and understanding of the structure of Nature at the smallest possible scales. The most recent and major milestone, the observation of the Higgs boson in 2012 at the LHC, was a tremendous confirmation of the SM. Indeed, No-Lose theorems based on precise mathematical principles that are believed to hold at the most fundamental level in the framework of QFT essentially guaranteed the existence of some form of new degrees of freedom needed to make mathematical sense of the theory. These guiding principles and their rigorous consequences provided clear justifications for a new accelerator after the Large Electron-Positron Collider (LEP). Furthermore, they motivated the conception of the LHC by pointing to a clear direction, in terms of an energy frontier, that would certainly lead to discoveries. The further exploitation of the LHC, however, did not yet deliver the discovery of NP which was expected by large parts of the HEP community.

The current situation and, more specifically, the future of our field in connection to the present status of our knowledge, is less transparent since there is no clear indication of a new energy frontier where NP may appear. All the particles of the SM have been discovered, and the SM is not only completely consistent at the mathematical level but also with almost all experimental measurements and tests. There are nevertheless clear signs and indications that NP exists, but they do not point, as was the case for the Higgs, to a specific energy range to probe. Neutrino oscillations imply non-zero mass differences among the three species, which is not accounted for in the SM. The observation of gravitational effects that could indicate the existence of Dark Matter (DM) and the observed matter-antimatter asymmetry in our Universe also suggest that the SM needs to be extended. From the point of view of experimental measurements, intriguing signals are provided by possible deviations from the SM prediction in the muon magnetic moment, the observation of a series of flavor anomalies in $B$ decays, and the recent tension in the $W$-boson mass measurements. Additionally, the determination of Higgs self-couplings, in connection to better understanding the shape of its potential, and Yukawa interactions offer further motivation to probe the SM with increased precision. Finally, theoretical puzzles and unexplained mathematical structures and free input parameters featured in the SM ask for an underlying yet-to-be-explored principle or mechanism. The smallness of the Higgs mass and the cosmological constant (two instances of \emph{hierarchy} problems), the highly hierarchical and peculiar pattern of masses and mixing angles among the three families of matter particles (the Flavor Puzzle), the absence of CP violation in the strong sector of the SM (the Strong CP problem), the quantization of $U(1)$ charges and the possible gauge couplings unification are all examples of theoretical puzzles and motivations to extend the SM. Ultimately, the unification of gravity with the three other fundamental forces is yet another important theoretical effort.

Nevertheless, the essential lesson to be learned from the current status in HEP is that the SM has to be viewed as \emph{an effective description} of physics valid only up to a certain energy scale. This energy scale, at which new and unknown degrees of freedom or dynamics become relevant, is not unambiguously indicated by any of the puzzles listed above, even though specific scenarios or measurements of some deviations can suggest a particular range of energies to be probed. 


The lack of knowledge about the next relevant energy scale does not, however, prevent the conception and design of a new collider. Indeed, multiple sectors of HEP would hugely benefit from new and more precise measurements which would improve upon the state-of-the-art and provide precious guidance toward NP. This new machine would certainly shed light on the current puzzles and issues listed above, reduce the allowed parameter space, or even exclude theoretical models, measure with improved precision the SM parameters and, in the best-case scenarios, provide clear evidence of an NP scale. Moreover, even in the absence of direct detection of new particles, precision measurements of scattering processes carry, via the quantum loop sensitivity inherent in QFT, information about heavier degrees of freedom and hence provide valuable input about the possible presence of NP.

The judicious selection of a specific new particle collider is thus of utmost importance for the next decades, as it will determine the energy scales and precision with which particular processes will be measured and, subsequently, which of the theoretical and experimental puzzles listed above will be addressed to which level of accuracy. Many different options exist, each offering particular scientific and technological challenges as well as societal and environmental impacts. Moreover, the collider selection will govern the future direction of ECR careers. 

\section{Requirements and Options for the next HEP machine}

The future collider options can be ordered by their shape (linear or circular) and the particles that are collided (leptons, hadrons, or lepton and hadrons). Leptons are fundamental particles, which lead to cleaner collisions compared to hadrons. On the other hand, leptons used in colliders (e.g. $e^-$, $e^+$) are generally lighter than hadrons (e.g. protons). They can therefore be accelerated more quickly to high momenta -- linear electron colliders are feasible while linear hadron colliders are not. On the other hand, the smaller mass of $e^-$/$e^+$ leads to higher beam energy losses due to synchrotron radiation when kept on a circular trajectory in a magnetic field. Therefore, circular electron-positron colliders can not achieve energies beyond the $ttH$ threshold, while linear electron-positron colliders can reach up to $3$ TeV with reasonable collider dimensions. Due to the recirculation of not collided particles, circular colliders achieve higher luminosities at lower $\sqrt{s}$. Additionally, they can easily support multiple detectors and interaction points, offering thereby a richer and broader physics program. Linear colliders are superior at larger $\sqrt{s}$ when circular colliders are limited by synchrotron radiation. Linear and circular $e^+ e^-$ colliders feature similar performance in the region around $240$ - $365$ GeV. Muon colliders have the same benefits as $e^+ e^-$ colliders, as muons are elementary particles as well, but the larger mass of the muon allows to build circular $\mu^+ \mu^-$ colliders that potentially could reach the TeV energy range as well.

In the following, the discussion will concentrate specifically on FCC-ee: this is not meant to express a preference in the choice\footnote{The reader is encouraged to check out references and resources related to other collider proposals in order to get a broader picture of the different projects for the future HEP machine.} of the future HEP machine, but to take advantage of the many resources and discussions at the workshop that particularly focused on FCC-ee as an example of future collider. Many of the following considerations also apply to other collider proposal, which are:

\begin{itemize}
    \item the International Linear Collider (ILC) \cite{ILCInternationalDevelopmentTeam:2022izu}, 
    \item the Cool Copper Collider (C$^3$) \cite{Bai:2021rdg},
    \item the Compact Linear Collider (CLIC) \cite{Brunner:2022usy},
    \item the Muon Collider \cite{Accettura:2023ked} and
    \item the Large Hadron Electron Collider (LHeC) / FCC-eh  \cite{Bruning:2022hro, LHeC:2020van}.
\end{itemize}


\subsection{FCC-ee}

The last update of the European Strategy for Particle Physics (ESPP \cite{European_Strategy_Update2020}) states that an electron-positron Higgs factory is the highest priority next collider. The Future Circular Collider (FCC \cite{fcc_physics}) at CERN is planned to be a versatile, next-generation particle collider housed in a 90.7 km underground ring aimed at combining a comprehensive battery of precision measurements with numerous opportunities for NP searches, at low and high energy. It will be implemented in two main stages: an $e^+e^-$ machine (FCC-ee \cite{fcc_ee}) followed by a high-energy hadron collider (FCC-hh \cite{fcc_hh}). The following paragraphs will describe the FCC-ee, but most considerations also apply to the Circular Electron-Positron Collider (CEPC \cite{cepc_tdr}).

FCC-ee is planned to be built between 2032 and 2045 and targets the intensity and precision frontier. Colliding electrons and positrons at energies between $90-365$ GeV will enable stress tests of the SM and a thorough exploration of indirect and low mass signals of Beyond the Standard Model (BSM) physics. The FCC-ee offers the highest luminosities of all proposed Higgs factories in a clean environment. It will provide an order-of-magnitude improvement in the measurements of several Higgs parameters with respect to the end of HL-LHC. In particular, sub-percent measurements of the Higgs couplings to $W$, $Z$, bottom quark, and $\tau$ and percent measurements of Higgs couplings to gluons and charm quarks are expected. A precise determination of the Higgs mass and width is achievable, together with the measurement of the Higgs self-couplings as one of the final goals of the complete FCC program. This will yield significant improvements in the understanding of the shape of the Higgs potential, relevant for the study of the vacuum structure and stability of the Universe. A run at the Higgs mass of 125 GeV would furthermore allow direct Higgs production at 125 GeV to measure the electron Yukawa coupling that gives mass to the electron. 

At $\sqrt{s} = 90$ and 160 GeV, FCC-ee is also an Electroweak (EW) factory. With unprecedented statistics (a factor of $\approx 3 \times 10^5$ better than at LEP1 and a factor of $\approx 5 \times 10^3$ better than at LEP2) it will provide comprehensive measurements of many Electroweak Precision Observables (EWPO), especially $\tau$-based EWPO, and an incredible improvement in the indirect sensitivity to BSM effects as well as precise estimations of $\alpha_{\text{QED}} (m_Z)$ and $\alpha_s(m_Z)$, at the per mille level. FCC-ee will also study, for the first time, top quarks in $e^+ e^-$ collisions, enabling the precise determination of the top quark mass (to an accuracy of $\pm 20$ MeV) and width.

The FCC-ee is also a remarkable flavor factory (in particular for $b$ and $\tau$ physics), providing a clean environment, precise energy and momentum resolutions, excellent vertexing, flavor tagging and hadron identification properties implemented via new Machine Learning (ML) techniques. It would offer interesting opportunities for the study of rare $B$ hadron decays with $\tau$-leptons in the final state and the exploration of flavor anomalies, lepton flavor violations and lepton-universality tests in $\tau$-decays. Since flavor physics is an important field to look for indirect signals of BSM physics and probe deep UV dynamics, it needs to be stressed that FCC-ee will compete with dedicated flavor experiments and provide, for example, $\sim 10$ times more bottom and charm pairs than Belle II. Moreover, it will be sensitive to flavor-violating operators generated by heavy new particles whose contributions are induced via Renormalization Group Running in un-flavored EWPO operators, an effect completely invisible at the LHC.

This collider will also enable access to the second-generation Yukawa couplings, specifically to $y_c$ and $y_s$, and potentially constrain $y_u$ and $y_d$. FCC-ee will also yield precise measurements of CKM entries, such as $|V_{cb}|$ and $|V_{ub}|$ as well as studies of rare $b$ and $c$ decays and CP violation in heavy quarks thanks to its excellent $c$ and $s$ tagging capabilities. Additionally, the precise measurements of Flavor Changing Neutral Current (FCNC) processes, which are extremely sensitive probes of BSM physics, and analyses of flavored jets will provide complementary information to meson data and significantly boost the synergies across these HEP programs.

FCC-ee will also provide an extremely interesting and broad array of physics measurements related to exotic BSM physics in connection to dark matter, neutrino masses or baryogenesis taking advantage of its foreseen detector hermiticity, tracking and calorimetry capabilities. In particular, it will offer a great environment for signature-driven searches of new unconventional and non-mainstream signatures related to WIMPs, ALPs, HNLs and FIPs as well as signals of dark sectors (e.g. dark photons, Higgs portals), exotic Higgs and $Z$ decays (e.g. to LLPs, featured in several BSM models) and light NP. These direct searches of (light) NP will probe parameter spaces of many BSM models that will not be constrained by astrophysical nor cosmological experiments, hence providing a complementary physics program to other HEP fields.

FCC-ee will not only offer an extraordinarily versatile tool to probe BSM physics but will also pave the way to the exploration of new, qualitatively different areas of HEP. Indeed, investigations of non-perturbative Quantum Chromodynamic (QCD) processes, quark-gluon fragmentation effects, and analyses of various structural features of QFT, such as spin correlations and entanglement, will be made possible. Interesting searches of non-decoupling NP will be possible at FCC-ee (via di-Higgs processes), and the parameter space of any model featuring non-decoupling BSM physics will be fully explored at this collider, making it an extremely powerful tool to probe unconventional and conceptually different areas of BSM. 

FCC-ee will offer an extremely rich and broad landscape of possible physics searches related to Higgs, flavor, QCD-EW, and (exotic) BSM physics, and feature many synergies among its stages and runs as well as with other HEP experiments (e.g. astrophysical, cosmological searches). It has strong complementary to the HL-LHC and great flexibility in terms of physics opportunities, thereby providing an incredible tool to maximize the potential for discoveries. The foreseen capabilities of this HEP machine will bring new exciting challenges in all areas of particle physics, hence requiring global progress in theoretical calculations (crucial for the calibration of the machine as well as the interpretation of the results), BSM model-building, software development, detector, and magnet technologies. Furthermore, a thorough study of the socio-economic ramifications and sustainability concerns associated with this new collider is essential to accurately quantify its effects. FCC-ee technology is mature, meaning that it can be built in parallel to HL-LHC operation, thereby providing a guarantee for the continuity of the HEP program, a crucial point for the ECR community in particular and for future HEP students, as well as for the community in general. The opportunity to enable the FCC-eh program will offer valuable complementary inputs, in particular on the structure of hadrons and on EWPO as well as on Higgs couplings measurements and BSM searches. A successful FCC-ee program will open the gate to the FCC-hh regime, which will benefit from the guidance and discoveries of the previous $e^+e^-$ program, thereby providing a clear and fruitful interplay among the two stages and a precise path for the future of high energy particle physics.

\subsection{General Requirements}
Regardless of the numerous options for the new HEP machine, several key features have to be satisfied by the new collider as they play an essential role in shaping the design and functionality of the new machine. 

An important aspect of the new particle collider is its inherent upgradability, the ability to easily enhance and modify its design to allow for continuous adaptation to evolving scientific requirements and technological advancements. The collider must be conceived as an adaptable system, avoiding a rigid design that could prevent the assimilation of new technologies. Such a design would allow for the gradual implementation of upgraded components and technologies over time, as they are being discovered and developed. This approach will facilitate the incorporation of state-of-the-art detectors, software, and electronics and ensure that the machine's capabilities can be continuously adapted and upgraded. The flexibility of a new HEP machine also plays an essential role during its development. Indeed, the long time scales required to design and build a new collider point to the necessity of constructing a machine that allows for the continuous incorporation of the latest technologies during its construction phase. In this context, planning and anticipating the development of new materials, magnets, software, experimental techniques, and other engineering features represents a key methodology that should be adopted. Finally, upgradability also extends the lifespan of the collider and guarantees that it remains at the forefront of experimental capabilities until its shutdown.

Another fundamental aspect is complementarity. The new machine must possess the versatility to allow for multiple different measurements and to address a wide and diverse spectrum of HEP experiments. It also has to be designed to complement and synergize with other experiments conducted throughout the world. Complementarity represents a key aspect in the current status of particle physics as there is no clear indication of any preferred region in the parameter space of new theories and extensions of the SM. NP could hide in unexpected corners of this space and it is hence of first importance to not sacrifice the exploration of any region without a clear theoretical motivation. Other experiments, already built or currently being designed, will feature specific sensitivities to particular energy scales and the new HEP collider must properly fit in the vast panorama of experiments conducted in the next decades to guarantee the diversity of the HEP program.

The High-Luminosity LHC program represents a significant enhancement in the performance of the LHC, set to be operational from the beginning of 2029. This upgrade aims to substantially increase the potential for discoveries of the existing collider. Bridging the gap between the High-Luminosity regime of the LHC and the capabilities of the new collider is another crucial requirement for the next HEP machine. This continuity ensures a coherent, thorough, and systematic exploration of the energy scales beyond the reach of the LHC. Options for the new HEP machine should ensure that this continuity is possible, both from the scientific point of view as well as from the perspective of construction time, available technologies, and possible knowledge transfer from the LHC to the new collider. This new machine has to provide interesting projects to bridge the gap with HL-LHC and should be designed in the most attractive and ECR career-compatible manner possible to limit the negative impact it could have on the ECR community.

\section{Challenges and objectives (Theory, precision calculations, software and computation)}
This section is aimed at answering question 5 and, more generally, the challenges that the different fields of HEP will have to face and address. The content of this section originates mainly from the discussions and talks presented at the one-day workshop at CERN, and as a result of the speakers' discussion points on the day, it is mostly FCC-centered.
\paragraph{Theory Opportunities:} From the theory perspective, many unresolved puzzles and questions are featured in the SM: the nature of Dark Matter, the matter-antimatter asymmetry, the peculiar flavor hierarchies, the absence of CP violation in QCD as well as fine-tuning problems. These issues and shortcomings do not unambiguously point towards a particular energy scale where NP contributions would address and solve these problems. Direct discovery of NP particles or dynamics would represent a tremendous achievement of the future collider, but would nevertheless appear, in the absence of clear and precise No Loose theorems and conjectures, as very fortuitous rather than the result of exact theoretical predictions. More reasonable objectives and requirements for a new HEP machine should be aimed at providing guaranteed deliverables (in terms of parameters and processes measured with some precision) as well as an unbiased and generic exploration potential for BSM physics. In the absence of clear theoretical guidance and specific motivation to restrict to a particular energy range or region in the parameter space of NP models, any new HEP machine should allow for enough flexibility to perform a wide range of different experiments and studies to efficiently and accurately scan any possible space where NP could hide.

The Higgs sector is a particularly well-motivated HEP direction in investigating BSM contributions. Indeed, there are good theoretical reasons to believe that the Higgs might be related to BSM physics which could shed light on e.g. the thermal history and the stability of the Universe, the origin of flavor and masses of the particles, CP violation, and baryogenesis. Moreover, the nature of the Higgs itself, also in connection to the Naturalness problems it introduces in the SM, is not completely clear, and this scalar particle could very well exhibit a more complicated inner structure (i.e. appear as a composite state instead of a fundamental particle) as well as portals to new hidden sectors of particle physics.

Depending on the type of the new HEP machine, different Higgs decay and production channels with different accuracy will be studied. Particularly interesting directions are the studies of Higgs couplings to second-generation fermions and the precise determination of the Higgs trilinear and quartic coupling, which are essential parameters closely related to the form of the Higgs potential and, ultimately, to the structure of the vacuum and stability of the Universe. Specifically for the investigation of the Higgs quartic, FCC-hh could provide an incredible factor 10 reduction in the measurement uncertainties compared to HL-LHC. FCC-ee on the other hand, while it would only improve the status on this measurement by a factor of 2, would still bring significant opportunities to test, at the sub-percent level, deviations in Higgs couplings compared to their SM predictions. This level of precision is crucial to assess many BSM models (e.g. 2HDM, Composite Higgs, some MSSM models) as these theoretical extensions predict modifications of a similar order of SM parameters. 

Other important Higgs-related observables, such as the Higgs mass and decay width, will provide another interesting physics case for a future collider. The large rates available at, for example, FCC-ee, together with the absence of pile-up and of QCD background as well as the precise knowledge of the center-of-mass energy offer an ideal environment for Higgs physics. It could allow for a model-independent measurement of the Higgs width of the order of $1\%$ and of the Higgs mass of the order of $\delta m_H \approx 3$ MeV, hence providing the EW fits with an extremely valuable and precise estimation of Higgs parameters. Measurements of Higgs couplings, in particular $H\rightarrow ZZ$, combined with measurements of $H\rightarrow XX$ at FCC-hh, will offer valuable synergies between the two programs, particularly for global fits. Moreover, this interplay will also benefit from percent-level precision in statistically limited rare channels (e.g. $H\rightarrow \mu \mu$) offered by FCC-hh, taking advantage of the large statistics in various Higgs decay modes. 

A Higgs factory would also investigate a large portion of the parameter space for axion-like and Heavy Neutral Leptons models, as well as various SM extensions featuring flavor non-universal gauge interactions and other flavor models. The absence of a clear energy scale at which NP may hide does certainly not imply that the community should restrain from working towards a new HEP machine, but it rather represents a formidable challenge to design an accelerator that could explore, in an unbiased manner, many regions of the parameter space and provide extremely accurate measurements of fundamental parameters of the SM.

\paragraph{Precision Calculation:} The foreseen precision capabilities as well as the enormous luminosity and event rates of the FCC-ee pose astounding and attractive challenges to theory predictions. This collider will allow for precise measurements of fundamental SM parameters at precision exceeding by two to three orders of magnitude those attained by the LHC, together with a drastic reduction of statistical (up to a factor of 500 compared to LHC) and systematic uncertainties. Improved theoretical calculations and predictions will thus be needed to fully exploit this new machine's discovery and exclusion potential, in particular, to enable robust comparisons between data and predictions to impose stringent constraints and bounds on possible NP contributions. Specifically, effective and accurate theory inputs are crucial for FCC-ee as they simultaneously enter into measurement/calibration (e.g. QED ISR) issues, in the interpretations of results, and parametric uncertainties of couplings and masses. This formidable task of providing predictions with extreme accuracy will be shared by all fields of precision calculations (fixed order and resummations in QCD and EW, EFTs, non-perturbative QCD, event generators, observables,...). Many technical challenges will have to be addressed, most of them requiring technologies yet to be developed.

FCC-ee will offer incredible opportunities both for electroweak and high-precision QCD physics. In the electroweak sector, despite the very clean environment of $e^+e^-$ collisions, knowledge and precise determination of higher-order effects will be crucial. Two- and three-loop EW corrections (e.g. for $Z\rightarrow qq+X$ or $e^+e^- \rightarrow e^+e^-$ for beam calibration), mixed QCD-EW effects, as well as initial and final state radiations, collinear radiations, QCD corrections to Higgs decay and precise estimations of hadronization and QED showers contributions, will be needed to take advantage of the machine's incredible capabilities at the $Z$-pole (improvement by three orders of magnitude compared to LHC) and of its promising potential as a Higgs factory (more than a factor of $10^5$ improvement on the number of produced $Z$-bosons compared to LEP while $\sigma_{ZH}$ is expected to be measured at $0.4\%$ precision). $W$-boson physics will also provide exciting possibilities, such as the determination of $\delta m_W \sim 5$ - $6$ MeV and the computation of $\mathcal{O}(\alpha \alpha_s)$ terms in $WW$ production.

High-precision QCD measurements will also benefit from the unique opportunities offered by FCC-ee. Precise measurements of $\alpha_s$ and studies of jet dynamics and substructures at high multiplicities and high perturbative order will pose great challenges to the development of theory calculations. At the level of precision offered by this new machine, understanding of $e^+e^-\rightarrow$ two, three or four jets at N$^3$LO and NNLO QCD, including mass effects, will become increasingly relevant, together with a study of Higgs form factors in $H\rightarrow$ jets, gluons and heavy quarks. Progress on PDF analysis will also be needed, as they are already a limiting factor at the LHC and HL-LHC. Remarkable breakthroughs will be required to investigate and model non-perturbative QCD effects, whose contributions might play an important role in the interpretation of several high-precision measurements (these effects can be of the order of a percent at the $Z$-pole), reflecting a major conceptual bottleneck for future improvements. Additionally, N$^3$LO QCD description of non-resonant contributions as well as a thorough assessment of off-shell effects will be needed to fully exploit the potential of the FCC-ee. In this context, it is also worth mentioning fruitful collaborations and active interest of the field of string theory and mathematical physics that started providing theoretical and mathematical frameworks in this perspective. Another important objective is to improve the accuracy and to develop new Monte Carlo (MC) parton shower generators to match the FCC-ee precision. Defining new observables is also a crucial point and LHC-gained expertise can be exploited. Finally, the incredible improvements offered by FCC-ee will not only impact the physics opportunities but also have direct consequences on technical requirements such as the extremely precise knowledge of the geometry and topology of the detectors, leading to the development of new technologies such as active laser alignment.

Progress on the theory side will be highly driven and encouraged by the current and foreseen LHC performances as well as the additional motivation provided by new colliders challenges and potential for discoveries. The synergies between the programs will provide fruitful developments and help bridge the gap between the two generations of colliders. In this context, and regarding the significant difficulties in improving the state-of-the-art of the field, large research and multi-disciplinary groups represent a key factor in allowing for independent cross-checks and fostering the development of new techniques. This strong effort in the design of new theory tools has to be combined with a deeper connection with the community of experimentalists as well as other fields of research (string theory, math community) and will consequently lead to an undeniable need for cutting-edge computational resources.

\paragraph{Software tools:} Computing infrastructures and software represent a key component of any particle physics program. Regarding the HL-LHC program and the development of a new HEP machine, the major challenge of designing more efficient computing models must be addressed by the community to meet the needs and fully exploit the opportunities offered by the new collider. A strong and common R\&D effort must be undertaken by the community in collaboration with other fields of science and industry to develop software and computing infrastructures that exploit recent technological advances and that will enhance particle collider experiments capabilities.

The need for better theoretical precision calculation (higher order QCD and EW corrections) together with the improvement of detector simulators and event generators will certainly lead to a higher demand for heavy computations and hence for a more efficient and accurate computational infrastructure. The main research directions in this respect are related to improving the efficiency and accelerating the expensive evaluation of matrix elements and phase-space samplings using High Performance Computing strategies which are increasingly GPU-based. Developing and enhancing the GPU prototypes and their capabilities, as well as improving event generation and detector simulation are crucial steps driven by the HL-LHC program and future colliders.

From a career perspective, computational physics needs to provide attractive opportunities for young researchers as maintenance and long-term software engineering strongly suffer from extremely heavy and time-consuming workloads along with poor recognition and visibility. Knowledge about software specifications and continuity in development and maintenance are key factors that need to be encouraged and supported to motivate young researchers to take responsibility and dedicate their careers to these essential aspects. Preserving and transferring knowledge as well as tooling for future experiments are indeed crucial, along with a close collaboration with software engineers and experimentalists to accurately specify the requirements and computing models as well as budgets needed for the new HEP machine and its experiments.

\section{Socio-economic impact and Sustainability}
This section discusses question 10 and presents a few of the most relevant aspects in connection to the societal and environmental impacts of future colliders.

Beyond purely scientific considerations, the project of building a new particle collider must also be studied from the perspective of the socio-economic outputs it will have on society. Its feasibility, cost-effectiveness, and territorial benefits, as well as the estimated costs and values it will provide, must be precisely assessed to quantify and distinguish how different options for a new HEP machine might affect society. Specifically in the FCC-ee study presented at the workshop, a model featuring six pathways through which the FCC-ee can generate socio-economic effects was considered, focusing exclusively on quantifiable and predictable costs and benefits that can be directly attributed to the FCC, from its construction to its final shutdown. The model aimed to provide a realistic estimate of the cost/benefit impact of the new machine taking into account the intrinsically probabilistic nature of such a long-term project by offering different perspectives and scenarios. Effects and benefits that were difficult or impossible to quantify and assess were dismissed such that the model provides a "minimum likely expected socio-economic effect" prediction on the project.

The impact of the scientific production of the project was studied from the perspective of the effects on the community based on the publications, papers, and conference proceedings of the FCC. The value of the production was quantified based on the citations and outputs it will generate in the scientific community, not by evaluating the novelty of its content nor the worth of the value presented.

From the point of view of training and human capital, the FCC will provide a rich and diverse work environment that will significantly affect, in a positive manner, the careers of its collaborators and members by improving their technical, scientific, and analytical skills. The gain at the labor market level was quantified by looking at the incremental salary received by CERN employees compared to peers who did not profit from the CERN experience. For CERN collaborators joining the industry or the private sector, the improvement was measured to be of the order of 3\% per year spent at CERN, mainly due to the added experience gained in the scientific sector and the development of problem-solving skills while working in the field of particle physics. This important aspect stresses the value of the future colliders as a human capital formation project that benefits its employees, collaborators, and the entire society. On the economic side, it was shown that every euro spent for the FCC would produce up to three euros of economic worth in industries and suppliers of the new HEP machine. The added value of building and operating the FCC was estimated to be of the order of $1.7$ billion CHF and 27000 jobs per year for the next 30 years were expected to be created. The value of companies and spin-offs created by CERN employees (averaging two new spin-offs per year) was also taken into account in the model.

The new machine will also bring a valuable cultural effect on society and offer, via outreach events and facilities, an extremely stimulating environment to popularize the field of HEP to the general public. Visitors, both on-site and virtually, also provide socio-economic value, either via the travel costs or the expenditure on-site. It was estimated that up to 600 CHF were spent onsite per visitor for a four-day visit on average, half of which directly benefits the local territories and economic network of France and Switzerland. More than 4 million incremental visitors were predicted to be expected as a result of FCC.

The model concluded that the measurable benefits of the FCC throughout its lifetime outperform the costs by 35\%, taking into account only the aforementioned measurable and predictable added values the new HEP machine will directly bring to society in terms of socio-economic worth. It should be stressed that the true value of the FCC, as seen by the general public and specifically by taxpayers, was investigated on a representative sample of the population in France, Switzerland, Germany, Israel, Italy, Poland, the UK, and the US. It was quantified based on the willingness to financially support the FCC-ee because of its perceived utility for humankind. The survey exhibits interesting variations depending on the countries, but the overall result is that the total public good value of the FCC-ee estimated via this method amounts to 570 billion CHF, significantly more than the 28.6 billion CHF of measurable benefits. This staggering difference illustrates the importance, as perceived by the population, of the FCC-ee as a tool to continue investigating and exploring the field of particle physics as well as expanding the boundaries of human knowledge. It certainly constitutes one of the strongest socio-economic arguments in favor of building a new HEP machine, as the financial support of the population for the FCC, based only on the public good and contribution it might provide to humankind, already vastly surpasses the measurable economical benefits it will bring to society.

Sustainability plays a particularly important role related to the future HEP machine. Its core concept is defined, according to the UN's 1997 report \cite{Brundtland}, as "development meeting present needs without compromising the needs of the future". It provides guidelines and promotes responsible actions, exemplified by the principle of reduce-reuse-recycle. CERN actively participated in creating the 17 UN sustainability goals and has committed to implementing these goals within its operations \cite{cernsust}.

One of the most urgent sustainability issues is certainly the massive emissions of Carbon Dioxide. $\text{CO}_2$ emissions are categorized into three scopes. Scope 1 emissions, originating directly from sources owned or controlled by CERN, are the dominant type of emissions and amount to almost 200000 tons of Carbon Dioxide Equivalent (t$\text{CO}_{2e}$) in 2018, mainly from cooling and operating detectors. Most of them are attributed to obsolete detector designs. Experiments have committed to achieving a 30\% reduction of these types of emissions after the Long Shutdown 3. Scope 2 emissions involve indirect emissions from energy procurement and usage (approx. 3000 t$\text{CO}_{2e}$ in 2018). Scope 3 emissions, which encompass activities indirectly related to CERN's operations, such as travel or food consumption, constitute less than 5\% of the organization's overall emissions.

From a more general perspective, CERN's commitment to sustainability is also illustrated by its ISO 50001 certification (the first laboratory to have it), an example of the importance given to sustainable practices by the management and the staff. This certification implies the establishment of improvement goals and continuous monitoring. CERN's sustainability efforts also extend into innovative technological and societal projects, such as the creation of an eco-district near CERN heated by the residual hot water from the cooling system of Point 8. This waste heat recovery and supply mechanism will provide heat to consumers in the vicinity at an attractive price. Moreover, the construction and operation of CERN and the FCC with renewable energy sources via long-term procured resources can be both economically appealing and provide, due to the allocation of overcapacities, electricity at an affordable price for other institutional and societal consumers. The treatment of excavation material also represents an important issue related to the future HEP machine, at the sustainability level as well as in connection to the socio-economic cost-benefit impact related to the transportation and deposit of this material. 

Additionally, the development of high-temperature superconductors and high-efficiency klystrons, crucial pieces of technology needed in accelerators, are further examples of the desire to improve energy efficiency and reduce the amount of carbon dioxide emissions. CERN is also committed to encouraging the transfer of new technologies to the public and is working in close collaboration with many companies on different sustainability-related projects, such as the study of future hybrid or electric planes (with Airbus), climate monitoring (with ESA) and the design of energy-efficient cooling and ventilation installation (with ABB) or large cryostats of liquid hydrogen (with GTT) \cite{ABB,Airbus,GTT}. 

In particular, FCC-ee would be the most efficient next-generation collider, in terms of energy consumption and carbon footprint per Higgs boson produced, due to its large luminosity \cite{Janot:2022jtn}. With four interaction points, the energy consumption per Higgs boson would be less than1.8 MWh, an order of magnitude less than ILC for example.  Nevertheless, ILC polarized beams offer exciting physics opportunities to access with high precision and efficiency EW physics. Additionally, as more than 90\% of CERN's electricity is produced from carbon-free (nuclear, hydro-electric,...) sources, the carbon cost of $1$ MWh would be drastically smaller than e.g. in the USA today. For the case of FCC, splitting the program into two stages (FCC-ee and FCC-hh) allows to spread the costs over a longer period and to reuse the same infrastructures, both aspects being beneficial from the point of view of their socio-economic as well as the sustainability impacts. The combined focus on minimizing the environmental footprint (CO2, energy, water, waste, resources) and on maximizing the physics output as well as the value returned to society, together with technological requirements and efficient design are important factors to be taken into account in order to have  the least disruptive HEP machine in terms of environmental impact during operation. \cite{Janot:2022jtn}.

Sustainability has evolved into an essential requirement and consideration for future projects. Integrating environmentally responsible practices, resource efficiency, optimization of technologies, and ethical/societal considerations without giving up on providing a solid and attractive physics case as well as a visible positive return to society (through education, outreach, and training) should be crucial requirements for the future HEP machine. A holistic approach is therefore required to make any new collider a driver of innovation and physics discoveries, well integrated into society with strong links to the industry, committed to minimizing environmental impact, and optimizing energy use to align with the global imperative of sustainability.

\section{Timescale and impact on ECR careers}
This section addresses questions 7, 9, and 11 and discusses the most relevant challenges ECR careers face in connection to future colliders.

The development of future particle colliders holds both promises and challenges for ECRs. Several inherent difficulties and constraints in the design and construction phase of the new HEP machine can significantly affect the careers of young physicists, and these issues should be very accurately described and addressed to minimize the negative impact on the ECRs.

The most substantial challenge is the extended timeline required for the design, validation, and construction of a new particle collider. ECRs are confronted with the necessity of committing to long-term projects such as detector development, accelerator, and software designs for the new machine that demand a significant amount of resources and time to be allocated. This poses a dilemma as specialization in areas that may not ultimately align with the needs of the new collider or absorb extensive time without tangible results can jeopardize the career trajectory of young researchers, especially in the context of short-term jobs and non-permanent positions. The risk of investing substantial time and expertise in a project that may not materialize as planned or that may undergo many modifications throughout its year-long development adds a layer of uncertainty that furthermore aggravates the decision-making in the context of career direction. The very long timeline is inherent to the development of a new machine and the whole field has to properly adapt to make the transition to the new generation of colliders more ECR career-friendly. This means, among other things, providing enough attractive projects and academically valuable initiatives to bridge the gap between the HL-LHC and the future machine such that the academic cost for the ECRs and the negative impact on their careers can be as minimal as possible.

In this context, it is essential to ensure the transferability of knowledge and skills from one experiment and project to another, both at the political level and at the technical level. R\&D projects related to future colliders must offer enough versatility to guarantee that the expertise and skills gained are transferable and valued across various initiatives and collaborations. This flexibility is crucial, allowing researchers to adapt and re-orient their careers if their initial specialization becomes obsolete or if the project takes an unforeseen turn. At the level of management, appreciation has to be given to expertise and skills developed rather than to specific knowledge or commitment to a particular project. The ability of ECRs to provide expertise across diverse and complementary areas is indispensable and should be encouraged, not only to prevent over-commitment to a single prototype or skill set but also to enhance the overall adaptability and flexibility required in this formidable task of building a new HEP machine.

ECRs should be encouraged to actively seek leadership roles and make their voices heard in decisions related to future colliders. This not only contributes to professional and academic growth but also ensures that the perspectives and opinions of young researchers are part of the decision-making and development of the new HEP machine.

\section{ECR impact on the choice of future colliders}
This section addresses question 12 and describes ways ECRs can make their voice heard on the choice of future colliders and related issues.

There are several ways that the ECR community, and in particular the ECR panel of ECFA, can give solicited and unsolicited input. Based on the mandate of the ECFA ECR panel, any issues of importance to the panel can be communicated to the wider public through ECFA's bi-annual newsletter, where the ECR panel always gets space to include its results and opinions. In addition, the ECR panel publishes an annual, slightly more detailed, report on arXiv \cite{panel2022ecfa}. 

Twice a year, Plenary ECFA (PECFA) meetings are taking place, in which the ECR panel has 5 delegates, and a slot in the agenda is always reserved, and used, for the ECR input to ECFA. Finally, within the mandate of the ECFA ECR panel, there is a representative who is an observer in the Restricted ECFA (RECFA) panel, who joins all meetings and country visits of RECFA. 

An example of solicited advice comes in the form of input to the strategy process of CERN on the decision of the next collider, for which a document from the ECR community is asked to be submitted.

Additionally to the solicited, and mandated, inputs to different bodies within ECFA, the ECR panel has the option to give unsolicited advice, write additional documents, or organise additional events or meetings when needed. One example of an additional event, with a corresponding document, is the future colliders meeting discussed in this document. As the timeline for the decision of the next collider was one of the major worries identified in the event, a letter was sent to the CERN council for its meeting in March 2024, an example of unsolicited advice. The letter is added in Appendix A.

\section{Conclusion}

This document summarised the discussions that took place at the \textit{Future Colliders for Early-Career Researchers} event that the ECFA ECR panel organized in September 2023. 

The future of particle physics is closely tied to the choice of the next-generation HEP machine. For the next few decades, this choice will not only dictate the main scientific direction as well as the physics program and discovery perspectives of our field, but also the technological, societal, and environmental effects and requirements it necessitates and implies. 

This document provided an overview of the main aspects in connection to different options for future colliders, from an ECR perspective. A particular emphasis has been given on the theoretical, computational, and technological implications and requirements needed to develop and fully exploit the potential for discovery. Socio-economical effects and sustainability issues focused on the FCC specifically have also been introduced and discussed.

Specifically for ECRs, the future HEP machine should emphasize the continuity of the physics program, to bridge the gap to HL-LHC and provide exciting project opportunities to younger scientists. Transferability of skills and knowledge, as well as expertise across the field, should be encouraged both at the scientific and political levels, in connection also to improving the current status of very short-term jobs and positions for young researchers. Moreover, an early, clear, and determined decision regarding which future machine will be built can only have positive effects on the ECRs community as well as on the whole field of particle physics as it will give a strong incentive to work towards a project that has high chances of being realized. Additionally, it can only positively affect the attractiveness of our field for the younger generation of physics students who will ultimately become scientists working on and using the future machine. We, therefore, believe that the highest priority and importance has to be given to the choice of the future HEP machine, that this choice has to be made as soon as possible in the most informed way to minimize the negative impacts on ECRs careers and maximize the physics and discovery opportunities, thereby providing a clear direction for the future of our field.

Since many aspects related to future colliders, such as funding and career opportunities, are country-dependent, national follow-up events are being organized. Such events have already taken place in some countries and the ECFA ECR panel is striving to have one such event in every ECFA member country, with the possibility of joining forces between neighboring countries. The ECFA ECR panel has therefore written a blueprint for such national \textit{Future Colliders for Early-Career Researchers} \cite{blueprint}.

In their March 2024 meeting, the CERN Council has decided on the timeline of the next European strategy update. Community input can be submitted until the end of March 2025. This process is scheduled to conclude in June 2026, when the Strategy is updated by the CERN Council. 

Due to the strong impact of the future collider decision on ECRs, ECRs should consider submitting a white paper input to the ESPP process to, among others, voice issues written in this document.


Finally, we want to thank all the speakers, poster presenters, and participants of the workshop at CERN for their engagement. We also thank the CERN Council and TH secretariats for their generous financial and organizational support for the event.

\section{Acknowledgments}

We are very grateful to the PECFA and RECFA in particular for their support in this event and the country-specific follow-up events mentioned in the document and continued support for ECR activities.
    
\bibliographystyle{atlasBibStyleWithTitle_50etAl}
\bibliography{main}

\providecommand{\href}[2]{#2}\begingroup\raggedright\begin{thebibliography}{10}

\bibitem{European_Strategy_Update2020}
{European Strategy Grou}p, \href{http://dx.doi.org/10.17181/ESU2020}{{\em {2020
  Update of the European Strategy for Particle Physics}},} tech. rep., Geneva,
  2020.

\bibitem{ILCInternationalDevelopmentTeam:2022izu}
{ILC International Development Team} Collaboration, A.~Aryshev {et~al.}, {\em
  {The International Linear Collider: Report to Snowmass 2021}},
  \href{http://arxiv.org/abs/2203.07622}{{\ttfamily arXiv:2203.07622
  [physics.acc-ph]}}.

\bibitem{Bai:2021rdg}
M.~Bai {et~al.}, {\em {C$^3$: A ''Cool'' Route to the Higgs Boson and Beyond}},
  in {\em {Snowmass 2021}}.
\newblock 10, 2021.
\newblock \href{http://arxiv.org/abs/2110.15800}{{\ttfamily arXiv:2110.15800
  [hep-ex]}}.

\bibitem{Brunner:2022usy}
O.~Brunner {et~al.}, {\em {The CLIC project}},
  \href{http://arxiv.org/abs/2203.09186}{{\ttfamily arXiv:2203.09186
  [physics.acc-ph]}}.

\bibitem{Accettura:2023ked}
C.~Accettura {et~al.}, {\em {Towards a muon collider}},
  \href{http://dx.doi.org/10.1140/epjc/s10052-023-11889-x}{Eur. Phys. J. C
  {\bfseries 83} (2023) 864}, \href{http://arxiv.org/abs/2303.08533}{{\ttfamily
  arXiv:2303.08533 [physics.acc-ph]}}, [Erratum: Eur.Phys.J.C 84, 36 (2024)].

\bibitem{Bruning:2022hro}
O.~Br\"uning, A.~Seryi,  and S.~Verd\'u-Andr\'es, {\em {Electron-Hadron
  Colliders: EIC, LHeC and FCC-eh}},
  \href{http://dx.doi.org/10.3389/fphy.2022.886473}{Front. in Phys. {\bfseries
  10} (2022) 886473}.

\bibitem{LHeC:2020van}
{LHeC, FCC-he Study Group} Collaboration, P.~Agostini {et~al.}, {\em {The Large
  Hadron\textendash{}Electron Collider at the HL-LHC}},
  \href{http://dx.doi.org/10.1088/1361-6471/abf3ba}{J. Phys. G {\bfseries 48}
  (2021) 110501}, \href{http://arxiv.org/abs/2007.14491}{{\ttfamily
  arXiv:2007.14491 [hep-ex]}}.

\bibitem{fcc_physics}
A.~Abada,  {et~al.}, {\em FCC Physics Opportunities: Future Circular Collider
  Conceptual Design Report Volume 1},
  \href{http://dx.doi.org/10.1140/epjc/s10052-019-6904-3}{The European Physical
  Journal C {\bfseries 79} (2019)},
  \url{http://dx.doi.org/10.1140/epjc/s10052-019-6904-3}.

\bibitem{fcc_ee}
A.~Abada,  {et~al.}, {\em FCC-ee: The Lepton Collider: Future Circular Collider
  Conceptual Design Report Volume 2},
  \href{http://dx.doi.org/10.1140/epjst/e2019-900045-4}{The European Physical
  Journal Special Topics {\bfseries 228} (2019) 261–623},
  \url{http://dx.doi.org/10.1140/epjst/e2019-900045-4}.

\bibitem{fcc_hh}
A.~Abada,  {et~al.}, {\em FCC-hh: The Hadron Collider: Future Circular Collider
  Conceptual Design Report Volume 3},
  \href{http://dx.doi.org/10.1140/epjst/e2019-900087-0}{The European Physical
  Journal Special Topics {\bfseries 228} (2019) 755–1107},
  \url{http://dx.doi.org/10.1140/epjst/e2019-900087-0}.

\bibitem{cepc_tdr}
{The CEPC Study Group}, {\em CEPC Technical Design Report -- Accelerator (v2)},
  2023.
\newblock \url{https://arxiv.org/abs/2312.14363}.

\bibitem{Brundtland}
B.~Report
  \url{https://www.are.admin.ch/are/en/home/media/publications/sustainable-development/brundtland-report.html}.

\bibitem{cernsust}
C.~K.~T. group
  \url{https://www.home.cern/news/news/knowledge-sharing/cern-publishes-its-environment-report-2021-2022}.

\bibitem{ABB}
C.~K.~T. group
  \url{https://home.cern/news/news/knowledge-sharing/abb-and-cern-identify-174-energy-saving-opportunity-laboratorys-cooling}.

\bibitem{Airbus}
C.~K.~T. group
  \url{https://home.cern/news/news/knowledge-sharing/cern-and-airbus-partnership-future-clean-aviation}.

\bibitem{GTT}
C.~K.~T. group
  \url{https://report2023-kt.web.cern.ch/applications/a-collaboration-between-gtt-and-cern-to-enable-liquid-hydrogen-maritime-transportation}.

\bibitem{Janot:2022jtn}
P.~Janot and A.~Blondel, {\em {The carbon footprint of proposed
  ${\mathrm{e}}^+{\mathrm{e}}^-$ Higgs factories}},
  \href{http://dx.doi.org/10.1140/epjp/s13360-022-03319-w}{Eur. Phys. J. Plus
  {\bfseries 137} (2022) 1122},
  \href{http://arxiv.org/abs/2208.10466}{{\ttfamily arXiv:2208.10466
  [hep-ph]}}.

\bibitem{panel2022ecfa}
E.~E.-C.~R. Panel,  {et~al.}, {\em The ECFA Early Career Researcher's Panel:
  composition, structure, and activities, 2021 -- 2022}, 2022.

\bibitem{blueprint}
A.~Ilg, {\em Blueprint for national events on future colliders for early-career
  researchers}, 2024.
\newblock \url{https://zenodo.org/doi/10.5281/zenodo.10869355}.

\end{thebibliography}\endgroup

\section*{Appendix A}
\label{sec:apendixa}
Dear CERN Council,

In the 70 years since its founding, CERN has not only established itself as the global centre of particle physics research but as a powerful symbol of international collaboration and scientific excellence. This would never have been possible without the unfaltering support offered by the CERN member states.

As a community, we feel immense pride and gratitude that we are part of this journey of scientific exploration and opportunity which CERN has pioneered. While the High-Luminosity LHC constitutes a much-anticipated and necessary advance in the LHC program, a clear path beyond it for our future in the field must be cemented with as little delay as possible. For the field to sustain the population, expertise, and enthusiasm required to overcome the challenges of what CERN’s next major project/accelerator will present, the ECR community needs certainty without delay that High Energy Physics has an immediate future beyond HL-LHC, and that funding and positions required to realise our future will grow rapidly. 

We, the ECFA Early-Career Researchers Panel, on behalf of the ECR community, would like to strongly urge the Council to make every effort to ensure that the process of evaluating, selecting and implementing potential future projects, which will define this century of High Energy Physics for Europe and the World, proceed with as quick a pace as possible, accelerating its time frame to start the European strategy process as early as possible and conclude by early 2026. This will go some way in helping further secure CERN’s unique position in science, technology and international cooperation for the next 70 years and beyond. 

Kind regards,

The ECFA Early-Career Researchers panel
\end{document}